\documentclass[]{aa}
\usepackage{times}
\usepackage{natbib}
\usepackage[dvips]{graphicx}
\usepackage{epsfig,longtable,lscape}
\usepackage{epsfig}
\usepackage{rotating}
\bibpunct[]{(}{)}{;}{a}{}{,}

\def\XMM{{\em XMM--Newton}}
\def\EPIC{{\em EPIC}}
\def\MOS{{\em MOS}}
\def\Mone{{\em MOS1}}
\def\Mtwo{{\em MOS2}}
\def\pn{{\em pn}}
\def\RGS{{\em RGS}}
\def\ROSAT{{\em ROSAT}}
\def\XTE{{\em RossiXTE}}

\def\RX{RX J0440.9+4431}

\def\TreA{{3A 0535+262}}

\begin{document}

\title{\XMM\ observation of the persistent Be/NS X--ray binary pulsar \RX}

\author{N. La Palombara\inst{1}, L. Sidoli\inst{1}, P. Esposito\inst{2}, A. Tiengo\inst{1,3}, S. Mereghetti\inst{1}}

\institute{INAF, Istituto di Astrofisica Spaziale e Fisica Cosmica - Milano, Via Bassini 15, I--20133, Milano, Italy
\and INAF, Osservatorio Astronomico di Cagliari, localit\`a Poggio dei Pini, strada 54, I--09012, Capoterra, Italy
\and IUSS-Istituto Universitario di Studi Superiori, viale Lungo Ticino Sforza 56, 27100 Pavia, Italy}

\titlerunning{\XMM\ observations of RX J0440.9+4431}

\authorrunning{La Palombara et al.}

\abstract{Many X--ray accreting pulsars have a soft excess below 10 keV. This feature has been detected also in faint sources and at low luminosity levels, suggesting that it is an ubiquitous phenomenon. In the case of the high luminosity pulsars ($L_{\rm X} > 10^{36}$ erg s$^{-1}$), the fit of this component with thermal emission models usually provides low temperatures ($kT <$ 0.5 keV) and large emission regions ($R \ge$ a few hundred km); for this reason, it is referred to as a `soft' excess. On the other hand, we recently found that in persistent, low--luminosity ($L_{\rm X} \sim 10^{34}$ erg s$^{-1}$) and long--period ($P >$ 100 s) Be accreting pulsars the observed excess can be modeled with a rather hot ($kT_{\rm BB} >$ 1 keV) blackbody component of small area ($R_{\rm BB} <$ 0.5 km), which can be interpreted as emission from the NS polar caps.
In this paper we present the results of a recent \XMM\ observation of the Galactic Be pulsar \RX, which is a poorly studied member of this class of sources.
We have found a best--fit period $P$ = 204.96 $\pm$ 0.02 s, which implies an average pulsar spin--down during the last 13 years, with $\dot P \simeq 6\times10^{-9}$ s s$^{-1}$. The estimated source luminosity is $L_{\rm X} \sim 8\times10^{34}$ erg s$^{-1}$: this value is higher by a factor $<$ 10 compared to those obtained in the first source observations, but almost two orders of magnitude lower than those measured during a few outbursts detected in the latest years. The source spectrum can be described with a power law plus blackbody model, with $kT_{\rm BB} = 1.34\pm0.04$ keV and $R_{\rm BB} = 273\pm16$ m, suggesting a polar--cap origin of this component.
Our results support the classification of \RX\ as a persistent Be/NS pulsar, and confirm that the hot blackbody spectral component is a common property of this class of sources.
}

\maketitle

\section{Introduction}\label{introduction}

Most of the X--ray binary pulsars (XBPs) are High Mass X--Ray Binaries (HMXRBs) in which a Neutron Star (NS) with magnetic field B $\sim 10^{12}$ G is accreting matter from a high--mass early--type star, either an OB supergiant or a Be star. They can be persistently bright, with luminosities in excess of 10$^{34}$ erg s$^{-1}$, or transient sources characterized by quiescent phases, with emission around 10$^{34}$ erg s$^{-1}$ or less, interrupted by bright outbursts reaching $L_{\rm X} \sim 10^{36-38}$ erg s$^{-1}$ \citep{Negueruela98,Reig07,Sidoli10}.

In these sources the X--ray spectra between 0.1 and 10 keV are usually described by a rather flat power--law, with photon index $\sim$ 1, but several XBPs have shown a marked \textit{`soft}' X--ray excess above the main power--law component \citep[see][ for a review]{LaPalombara&Mereghetti06}; it is well described by a thermal emission model (either blackbody, bremsstrahlung or mekal) with low temperature ($kT_{\rm SE} <$ 0.5 keV) and large emission area ($R_{\rm SE} \ge$ a few hundred km). This feature has been detected not only in the high--luminosity sources (with $L_{\rm X} \sim 10^{37-38}$ erg s$^{-1}$) but also in several low--luminosity ($L_{\rm X} \sim 10^{35-36}$ erg s$^{-1}$) XBPs observed in the Small Magellanic Cloud (SMC), where its detection is favoured by the low interstellar absorption \citep{Sasaki+03,Ueno+04,Majid+04,Haberl&Pietsch05,Haberl+08}. Only in a few cases this low--energy component showed coherent pulses and the debate over its origin remains open. \citet{Hickox+04} have shown that a soft spectral component could be a very common, if not ubiquitous, feature intrinsic to X--ray pulsars: it is visible in all sources with a sufficiently high flux and small absorption, and its origin is related to the source total luminosity.

Recently, based on \XMM\ data, we have observed a clear thermal excess also in three of the four \textit{persistent} Be pulsars originally identified by \citet{Reig&Roche99}, i.e. \mbox{RX J0146.9+6121/LS I +61$^{\circ}$ 235} \citep{LaPalombara&Mereghetti06}, \mbox{4U 0352+309/X Persei} \citep{LaPalombara&Mereghetti07}, and \mbox{RX J1037.5-5647/LS 1698} \citep{LaPalombara+09}. These three sources are characterized by a persistently low luminosity ($L_{\rm X} \sim 10^{34-35}$ erg s$^{-1}$) and a long pulse period ($P >$ 100 s). These properties suggest that the NS orbits the Be star in a wide and nearly circular orbit, continuously accreting material from the low--density outer regions of the circumstellar envelope; in the case of 4U 0352+309 this picture is supported by the long orbital period of 250.3 days \citep{Delgado-Marti+01}. For these sources the detection of the thermal component was favoured by the small distance ($d \le$ 5 kpc) and interstellar absorption ($N_{\rm H} \sim 10^{21}$ cm$^{-2}$). We found that their soft excess can be fitted only with a blackbody (other simple models are rejected), which contributes for 30-40 \% of the total flux; interestingly, in comparison with the other, more luminous sources, their blackbody component is characterized by a higher temperature ($kT_{\rm BB} >$ 1 keV) and a much smaller emission radius ($R_{\rm BB} <$ 0.5 km). This \textit{hot BB} spectral component sets these low--luminosity and long--period sources apart from all the other pulsars, strongly suggesting that they form a distinct class. In their case the thermal component could be due to a different emission mechanism than in the high--luminosity pulsars. Based on the work of \citet{Hickox+04}, it can be attributed to emission from the neutron--star polar caps. This is supported by the emission area of the blackbody component, which is consistent with the estimated polar cap size, and by the fact that the low energy part of the spectrum is clearly pulsed.

In this paper we present the results of a recent \XMM\ observation of \RX, the remaining member of this class of Be/NS pulsars. This system was discovered during the \ROSAT\ Galactic plane survey \citep{Motch+97} and identified with LS V +44 17, a moderately reddened ($E(B-V)=0.65\pm0.0$5) B0.2 Ve star at $\sim$ 3.3 kpc \citep{Reig11}. Thanks to observations with the PCA instrument on board \XTE, \citet{Reig&Roche99} performed the first detailed timing and spectral analysis, and discovered a pulsation with period $P = 202.5 \pm 0.5$ s. Its spectrum was well fitted with different models (power--law, power--law plus blackbody, two black--bodies, cut--off power--law) and the measured flux implied a source luminosity of 3$\times$10$^{34}$ erg s$^{-1}$ between 3 and 30 keV. \RX\ has been detected also in the hard X--ray range: it is reported (as source PBC J0440.9+4432) in the Palermo \textit{Swift}-BAT hard X--ray catalogue \citep{Cusumano+10}, with a 15-150 keV flux of $(2.0\pm1.1)\times10^{-11}$ erg cm$^{-2}$ s$^{-1}$, and in the catalogue obtained with the \textit{INTEGRAL/IBIS} 7--year All--Sky Hard X--ray Survey \citep{Krivonos+10a}, with a 17-60 keV flux of $(1.36\pm0.22)\times10^{-11}$ erg cm$^{-2}$ s$^{-1}$.

In the optical/IR waveband, the H$\alpha$ line shows a double--peak profile, varying from symmetric shape to completely distorted on one side (V/R phases), with a correlation between the equivalent width of the H$\alpha$ line and the infrared magnitudes: as the EW(H$\alpha$) decreases the IR magnitudes become fainter. This long--term optical/IR variability is attributed to structural changes in the Be star's circumstellar disc, which alternates decline and recovery phases on typical time scales $T_{\rm disc} >$ 10 yrs \citep{Reig+05}.

During the \XTE\ observation this pulsar showed little X--ray variability (by a factor $<$ 10), therefore it was included in the class of persistent, low--luminosity and long--period Be pulsars. However, recently this source has shown three consecutive flux increases, spaced by $\sim$ 5 months from each other: the first one was detected by \textit{MAXI/GSC} on 31 March 2010 \citep{Morii+10} and by \XTE\ on 6 April 2010 \citep{Finger&Camero-Arranz10}; the second one was observed by \textit{INTEGRAL} on 1 September 2010 \citep{Krivonos+10b}; the third one was seen by \textit{Swift} on 29 January 2011 \citep{Tsygankov+11}. The mean time between the starts of the outbursts was $\sim$ 155 days, and the peak luminosities of these flares were lower than 10$^{37}$ erg s$^{-1}$ (assuming a source distance of 3.3 kpc). Assuming that these events occured at the periastron passage of the NS, where it approaches the decretion disc of the Be star, their separation is a good estimate of the orbital period of the system. It is interesting to note that the value of 155 days is also in agreement with the period of about 150 days derived from the Corbet diagram of $P_{spin}$ vs $P_{orbit}$ \citep{Corbet86}. Therefore, after 4U 0352+308/X Persei, \RX\ could be the second persistent Be/X-ray pulsar with a known orbital period.

\section{Observations and data reduction}\label{data}

\RX~was observed with \XMM\ on 2011 March 18 (MJD = 55638.417). The three \EPIC~cameras, i.e. one \pn\ \citep{Struder+01} and two \MOS\ \citep{Turner+01}, were operated in \textit{Large Window} mode, with a time resolution of 48 ms for the \pn\ camera and of 0.9 s for the two \MOS\ cameras; the effective source exposure time was, respectively, of $\sim$ 14 ks and $\sim$ 17 ks. For all cameras the medium thickness filter was used. \RX\ was also observed for $\sim$ 17 ks by the \textit{Reflection Grating Spectrometer} (\RGS), which was operated in \textit{Spectroscopy} mode \citep{denHerder+01}.

We used  version 11.0 of the \XMM~{\em Science Analysis System} (\texttt{SAS}) to process the event files. After the standard pipeline processing, we looked for possible intervals of high instrumental background, with negative result. \EPIC\ source events were selected within a circular area, with an extraction radius of 30$''$ for all the cameras; the corresponding background events were accumulated on large circular areas free of sources and with radii of 120$''$, 50$''$  and 60$''$ for the \pn, \Mone, and  \Mtwo\ cameras, respectively. We selected all the events in the energy range 0.15--12 keV, with pattern range 0--4 (i.e. mono-- and bi--pixel events) for the \pn\ camera and 0--12 (i.e. from 1 to 4 pixel events) for the two \MOS. In all cases the background contribution to the total \textit{count rate} (CR) was negligible, resulting in a net CR of $\sim$ 5.8 cts s$^{-1}$ for the \pn\ and $\sim$ 1.8 cts s$^{-1}$ for each of the two \MOS. Although the source CR was very high, we checked with the \texttt{SAS} task \textit{epatplot} that no event pile--up affected our data.

\section{Timing analysis}\label{timing}

\begin{figure}[h]
\centering
\resizebox{\hsize}{!}{\includegraphics[angle=-90,clip=true]{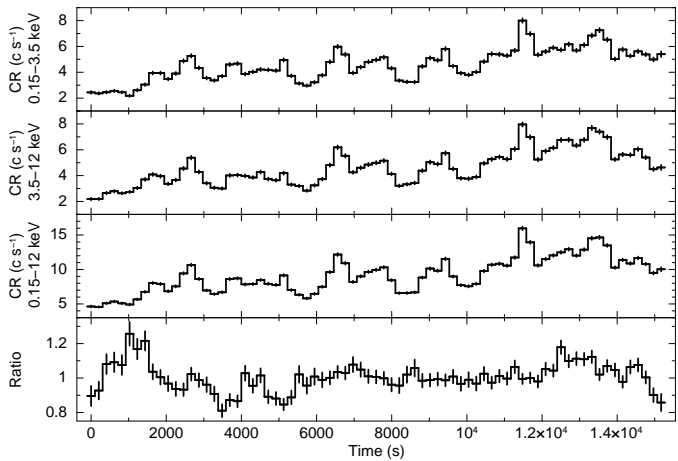}}
\caption{Background--subtracted light curves of \RX~in the energy ranges 0.15--3.5, 3.5--12, and 0.15--12 keV, with a time bin of 205 s (i.e. one pulse period).}
\label{lc_1P}
\end{figure}

For the timing analysis we considered only the \EPIC\ data. To measure the pulse period, we converted the event arrival times to the solar system barycenter and combined the three datasets. We measured the pulse period by a standard phase--fitting technique \citep{dallosso+03}, obtaining a best--fit period \textit{P} = 204.96 $\pm$ 0.02 s. This result takes into account the effects of the source variability during the observation (Fig.~\ref{lc_1P}); in fact, the individual pulses are detected with high significance along the whole observation, even when the source is faint.

\begin{figure}[h]
\centering
\resizebox{\hsize}{!}{\includegraphics[angle=-90,clip=true]{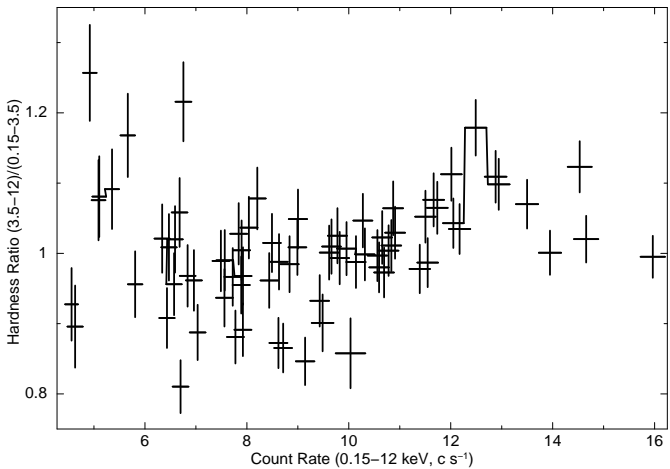}}
\caption{Hardness--ratio variation of \RX~as a function of the 0.15--12 keV count rate, with a time bin of 205 s.}
\label{hr_cr_sum}
\end{figure}

In Fig.~\ref{lc_1P} we report the background--subtracted light curves in the energy ranges 0.15--3.5 (soft), 3.5--12 (hard) and 0.15--12 keV (total), together with the \textit{hardness--ratio} HR between the hard (H) and soft (S) light curves (computed as H/S); the two energy ranges were defined in order to obtain a comparable number of counts, while the time bin of 205 s, corresponding to one pulse period, was chosen to avoid the effects due to the periodic pulsations. In both ranges the average CR is $\sim$ 4.5 cts s$^{-1}$, with an increasing trend along the observation; moreover, it was highly variable over the short time--scale, since there are CR variations up to $\sim$ 30 \% between consecutive time bins. Also the HR is characterized by a similar variability but, in this case, no long time--scale trend is observed; there is also no clear correlation with the source CR (Fig.~\ref{hr_cr_sum}), although the fit with a constant value HR = 1 is rejected ($\chi^{2}_{\nu}$/d.o.f. = 3.62/74).

\begin{figure}[h]
\centering
\resizebox{\hsize}{!}{\includegraphics[angle=-90,clip=true]{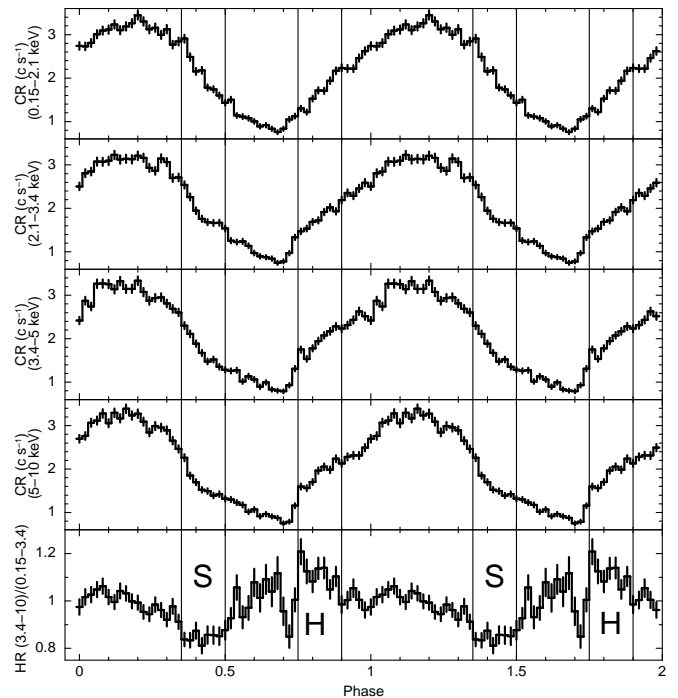}}
\caption{Background--subtracted light curves of \RX~in the energy ranges 0.15--2.1, 2.1-3.4, 3.4--5, and 5--10 keV, folded at the best--fit period \textit{P} = 204.96 s.}
\label{flc_4E}
\end{figure}

The \XMM\ observation allowed us to investigate the pulse profile, for the first time, even at energies below 3 keV. In Fig.~\ref{flc_4E} we show the folded light curves in 
four different energy ranges, defined in order to obtain a comparable number of counts in each of them, together with the hardness ratio between the 3.4--10 and 0.15--3.4 keV energy ranges. The shape of the pulse profile is similar in the four ranges: in all cases it shows a single broad peak, but it is not possible to fit it with a simple sinusoidal model; the measured pulsed fraction, defined as (CR$_{\rm max}$ -- CR$_{\rm min}$)/(2 $\times$ CR$_{\rm average}$), is $\sim$ 55 \% for all the energy ranges; the increasing part of the curve is more regular than the decreasing one, which is steeper at the beginning and flatter at the end. However, we note an energy dependence of the pulse profile around the CR minimum (phase $\phi$ = 0.7): the minimum CR value is reached at slightly later phases for increasing energies, and also the following CR increase is delayed; moreover, at the minimum egress the CR increases more suddenly at high energies than at the low ones. As a consequence, the HR between the hard and soft curves is highly variable in this phase interval: it first shows a sharp minimum just after the CR minimum, suddenly followed by a large peak.

\section{Spectral analysis}\label{spectra}

For the \EPIC\ source and background spectra, we adopted the same extraction parameters used for the light curves; we generated the applicable response matrices and ancillary files using the \texttt{SAS} tasks \texttt{rmfgen} and \texttt{arfgen}. We considered also the \RGS\ data, using the source and background spectra and the response matrices obtained with the standard reduction pipeline. In order to ensure the applicability of the $\chi^{2}$ statistics, the \EPIC\ and \RGS\ spectra were rebinned with a minimum of 100 and 30 counts per bin, respectively; then they were fitted in the energy range 0.3--12 keV using \texttt{XSPEC} 12.4.0. In the following, all spectral uncertainties and upper limits are given at 90 \% confidence level for one interesting parameter, and we assume a source distance of 3.3 kpc.

After checking that separate fits of the three \EPIC\ cameras gave consistent results, we fitted them simultaneously to increase the statistics; then we checked that both \RGS\ spectra were consistent with the \EPIC\ ones, therefore we included also them in the spectral analysis; to this aim, we introduced relative normalization factors among the spectra of the five cameras. Using an absorbed power--law (\textit{PL}) model, we obtained a hydrogen column density $N_{\rm H} = (1.19\pm0.02)\times 10^{22}$ cm$^{-2}$ and a photon--index $\Gamma$ = 1.16$\pm$0.01, with $\chi^{2}_{\nu}$/d.o.f. = 2.097/1083. Using an absorbed blackbody (\textit{BB}) model, we obtained $N_{\rm H} = (3.33\pm0.0.9)\times 10^{21}$ cm$^{-2}$, a \textit{BB} temperature $kT_{\rm BB} = 1.58 \pm 0.01$ keV, and a \textit{BB} radius $R_{\rm BB} = 306 \pm 3$ m, with $\chi^{2}_{\nu}$/d.o.f. = 2.604/1083.

\begin{figure}[h]
\centering
\resizebox{\hsize}{!}{\includegraphics[angle=-90,clip=true]{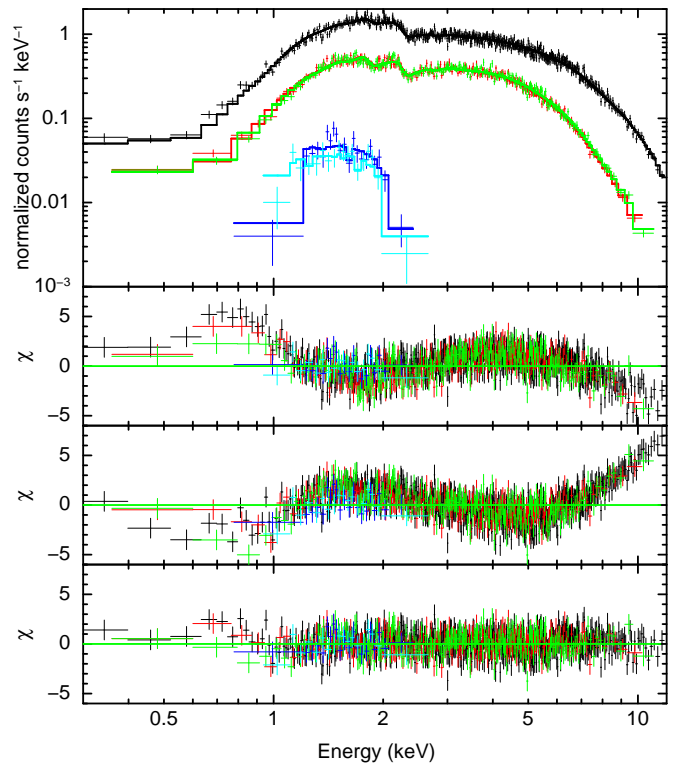}}
\caption{\textit{Top panel}: total spectrum of \RX~with the best--fit \textit{PL+BB} model. The spectra of the \pn, \Mone\ and \Mtwo\ cameras are shown in black, red, and green, respectively, while those of the \textit{RGS1} and \textit{RGS2} are shown in blue and light blue. \textit{Middle panels}: residuals (in units of $\sigma$) between data and model in the case of the single \textit{PL} and of the single \textit{BB}. \textit{Bottom panel}: residuals in the case of the \textit{PL+BB}.}\label{all_spetrum_powbb}
\end{figure}

In both cases the fits of the spectrum were unacceptable, with large residuals, so we repeated the fit with a \textit{PL+BB} model. In this way we obtained a significant improvement (Fig.~\ref{all_spetrum_powbb}), with $\chi^{2}_{\nu}$/d.o.f = 1.105/1081. The corresponding best--fit parameters are $N_{\rm H} = (7.3\pm0.4) \times 10^{21}$ cm$^{-2}$, $\Gamma$ = 0.85 $\pm$ 0.07, and $kT_{\rm BB}$ = 1.34 $\pm$ 0.04 keV. The \textit{PL} normalization is $I_{\rm PL} = (1.8 \pm 0.2)\times10^{-3}$ ph cm$^{-2}$ s$^{-1}$ keV$^{-1}$ at 1 keV and the \textit{BB} radius $R_{\rm BB}$ = 273 $\pm$ 16 m. The absorbed flux in the energy range 0.3--12 keV is $f_{\rm abs,X} = 6.0^{+0.2}_{-0.3} \times10^{-11}$ erg cm$^{-2}$ s$^{-1}$, while the corresponding unabsorbed flux is $f_{\rm unabs,X} = 6.7 \times10^{-11}$ erg cm$^{-2}$ s$^{-1}$, which implies a source luminosity $L_{\rm X} = 8.3\times10^{34}$ erg s$^{-1}$. The \textit{BB} component contributes for about the 35 \% to the 0.3-12 keV source flux.

Neither the \EPIC\ nor the \RGS\ spectra show clear emission or absorption features. We also looked for narrow iron K$_{\alpha}$ emission lines between 6 and 7 keV, with different widths between 0 and 0.5 keV. We found no evidence of such a component, with an upper limit on its equivalent width of $\sim$ 70 eV (at 90 \% c.l.) at most.

\section{Phase--resolved spectroscopy}\label{spectroscopy}

As shown in Fig.~\ref{flc_4E}, the HR is characterized by large variations along the pulse: therefore it is interesting to investigate the spin--phase resolved spectrum. To this aim, we first analyzed the background subtracted spectra of the \EPIC\ data in two different phase intervals, i.e. the \textit{soft} (S, $\phi$ = 0.35--0.50) and the \textit{hard} (H, $\phi$ = 0.75--0.90): in spite of the HR difference, they are characterized by a comparable CR in the \textit{total} (0.15--12) energy range and, then, of photon counts in the accumulated spectra.

The independent fits of the two spectra with a single \textit{PL} or \textit{BB} model do not provide satisfactory results (in some cases they are rejected by the data), while the use of a \textit{PL+BB} model improves significantly the spectral fit goodness; the best--fit parameters are reported in Table~\ref{2spectra_separate}. Taking into account the estimated uncertainties, none of the spectral parameters shows a significant variability between the S and the H spectra: in all cases they are characterized by comparable or consistent best--fit values.

\begin{table}[htbp]
\caption{Best--fit spectral parameters for the phase--resolved spectroscopy of \RX, in the case of the independent fit of the two spectra.}\label{2spectra_separate}
\begin{center}
\begin{tabular}{c|cc} \hline
Phase interval				&		S			&	H				\\ \hline
$N_{\rm H}$ ($\times10^{22}$ cm$^{-2}$)	& 0.76$^{+0.13}_{-0.14}$		& 0.69$^{+0.16}_{-0.17}$		\\ \hline
$\Gamma$				& 1.12$^{+0.26}_{-0.24}$		& 0.89$^{+0.35}_{-0.42}$		\\
$f_{\rm PL}^{a}$			& 2.76$^{+0.08}_{-0.31}$ (63 \%)	& 2.28$^{+0.19}_{-0.30}$ (44 \%)	\\ \hline
$kT_{\rm BB}$ (keV)			& 1.35$^{+0.17}_{-0.13}$		& 1.54$^{+0.12}_{-0.11}$		\\
$R_{\rm BB}$ (m)			& 237$^{+49}_{-37}$			& 247$^{+36}_{-27}$			\\
$f_{\rm BB}^{a}$			& 1.63$^{+0.25}_{-0.05}$ (37 \%)	& 2.95$^{+0.91}_{-0.15}$ (56 \%)	\\ \hline
$f_{\rm TOT}^{a}$			& 4.39$^{+0.18}_{-0.42}$		& 5.23$^{+0.22}_{-1.25}$		\\ 
$\chi^{2}_{\nu}$/d.o.f.			& 1.15/153				& 1.09/170				\\ \hline
\end{tabular}
\end{center}
\begin{small}
$^{a}$ Absorbed flux in the energy range 0.3--12 keV, in units of $10^{-11}$ erg cm$^{-2}$ s$^{-1}$
\end{small}
\end{table}

We also investigated the spectral variability at the pulse minimum, where the HR shows an abrupt variation, by considering the spectrum of two narrow phase ranges at $\phi$ = 0.72 and $\phi$ = 0.76. Due to the limited count statistics it was not possible to perform a detailed spectral analysis, but we checked that they are consistent with the best--fit models of the S and H phase intervals, respectively.

To investigate the relative variations of the two components with the period phase, we also simultaneously fitted the S and H spectra with the \textit{PL+BB} model forcing the same value of $N_{\rm H}$, $\Gamma$, and $kT_{\rm BB}$ for the two phase intervals. In this case we obtained $N_{\rm H}$ = ($0.92^{+0.08}_{-0.07}$)$\times10^{22}$ cm$^{-2}$, $\Gamma_{\rm PL}=1.48^{+0.17}_{-0.13}$, and $kT_{\rm BB} = 1.67 \pm 0.05$ keV, with $\chi^{2}_{\nu}$/d.o.f. = 1.17/328. The corresponding normalization values are reported in Table~\ref{2spectra_common} (\textit{Case 1}). With these constraints on the model parameters, the spectral changes as a function of the phase are reproduced by the variations in the relative contribution of the two components. The values reported in Table~\ref{2spectra_common} show that, with this assumption, the observed spectral variability is due to both components, since their flux varies of more than 50 \% between the two phases, even if in opposite directions: the \textit{PL} flux decreases and the \textit{BB} flux increases from the S to the H phase.

In order to confirm that the thermal component varies as a function of the rotational phase, we should prove that a constant \textit{BB} component is rejected by the data. To this aim, we modified the test model by linking the \textit{BB} parameters for the two spectra together, while both the photon--index $\Gamma$ and the normalization $I_{\rm PL}$ of the \textit{PL} component could vary independently in the two phase intervals. The resulting best fit gives $N_{\rm H}= (0.84 \pm 0.09)\times10^{22}$ cm$^{-2}$, $kT_{\rm BB} = 1.54 \pm 0.08$ keV, and $R_{\rm BB}$ = $223^{+17}_{-16}$ m. The power--law parameters are shown in Table~\ref{2spectra_common} (\textit{Case 2}). Even with this model we found a good fit quality ($\chi^{2}_{\nu}$/d.o.f. = 1.16/328), fully comparable to the previous case. This result suggests that it is possible to attribute the whole spectral variability to the \textit{PL} component, and that a constant \textit{BB} cannot be ruled out.

\begin{table*}[htbp]
\caption{Best--fit values for the black--body and power--law parameters, when the two spectra are fitted simultaneously with common values of $N_{\rm H}$, $\Gamma$, and $kT_{\rm BB}$ (\textit{case 1}) or with common values of $N_{\rm H}$, $kT_{\rm BB}$, and $R_{\rm BB}$ (\textit{case 2}).}\label{2spectra_common}
\begin{tabular}{c|cc|cc} \hline
Phase	 		&		\multicolumn{2}{c}{case 1}					&		\multicolumn{2}{c}{case 2}				\\
interval 		&		S 			&		H			&	S				&	H			\\ \hline
$N_{\rm H}^{a}$		& \textit{0.92$^{+0.08}_{-0.07}$}	& \textit{0.92$^{+0.08}_{-0.07}$}	& \textit{0.84 $\pm$ 0.09}		& \textit{0.84 $\pm$ 0.09}	\\ \hline
$\Gamma$		& \textit{1.48$^{+0.17}_{-0.13}$}	& \textit{1.48$^{+0.17}_{-0.13}$}	& 1.52$^{+0.22}_{-0.19}$		& 1.00$^{+0.16}_{-0.15}$	\\
$I_{\rm PL}^{b}$	& 3.37$^{+0.55}_{-0.46}$		& 2.24$^{+0.49}_{-041}$			& 2.65$^{+0.68}_{-0.59}$		& 1.83$^{+0.44}_{-0.40}$	\\
$f_{\rm PL}^{c}$	& 2.34$^{+0.28}_{-0.42}$ (54 \%)	& 1.56$^{+0.09}_{-0.22}$ (31 \%)	& 1.73$^{+0.10}_{-0.28}$ (42 \%)	& 2.93$^{+0.06}_{-0.26}$ (55 \%)\\ \hline
$kT_{\rm BB}$		& \textit{1.67 $\pm$ 0.05}		& \textit{1.67 $\pm$ 0.05}		& \textit{1.54 $\pm$ 0.08}		& \textit{1.54 $\pm$ 0.08}	\\
$R_{\rm BB}$		& 176$^{+13}_{-14}$			& 231$^{+12}_{-13}$			& \textit{223$^{+17}_{-16}$}		& \textit{223$^{+17}_{-16}$}	\\
$f_{\rm BB}^{a}$	& 2.02$^{+0.07}_{-0.10}$ (46 \%)	& 3.47$^{+0.05}_{-0.06}$ (69 \%)	& 2.42$^{+0.06}_{-0.10}$ (58 \%)	& 2.42$^{+0.06}_{-0.10}$ (45 \%)\\ \hline
$f_{\rm TOT}^{a}$	& 4.36$^{+0.10}_{-0.16}$		& 5.03$^{+0.14}_{-0.11}$		& 4.15$^{+0.12}_{-0.20}$		& 5.35$^{+0.20}_{-0.28}$	\\
$\chi^{2}_{\nu}$	& 	\multicolumn{2}{c}{1.17}						& 	\multicolumn{2}{c}{1.16}					\\
d.o.f.			& 	\multicolumn{2}{c}{328}							& 	\multicolumn{2}{c}{328}						\\ \hline
\end{tabular}
\begin{small}
\\
$^{a}$ $\times10^{22}$ cm$^{-2}$
\\
$^{b}$ $\times10^{-3}$ ph cm$^{-2}$ s$^{-1}$ keV$^{-1}$ at 1 keV
\\
$^{c}$ Absorbed flux in the energy range 0.3--12 keV, in units of $10^{-11}$ erg cm$^{-2}$ s$^{-1}$
\end{small}
\end{table*}

\section{Discussion}\label{sec:6}

The \XMM\ observation of \RX\ is the first one of this source performed with a large X--ray telescope of the last generation; it has allowed us to investigate, for the first time, the timing and spectral properties of this pulsar also at low X--ray energies below 3 keV.

We have obtained a new measurement of the pulse period: \textit{P} = 204.96 $\pm$ 0.02 s. This value is in agreement with the period measurements obtained by the Gamma--ray Burst Monitor on--board the \textit{FERMI} satellite since its launch in 2008\footnote{http://gammaray.nsstc.nasa.gov/gbm/science/pulsars/lightcurves/rxj0440\_fig1.png}: they show that this source is characterized by a variable spin period, which is typical of wind--fed binary systems. On the longer time--scale, our value is slightly larger than the previous value of 202.5 $\pm$ 0.5 s found in 1998 by \XTE\ \citep{Reig&Roche99}: the period difference ($\Delta P/P \simeq 1 \%$) is too large to be ascribed to orbital motion and implies an average spin--down during the past 13 years, with average $\dot P = (6.4 \pm 1.3)\times10^{-9}$ s s$^{-1}$.

We have estimated a source luminosity $L_{\rm X} \sim 6 \times 10^{34}$ erg s$^{-1}$ in the 2--10 keV energy range. To investigate the long--term evolution of \RX, we considered also the X--ray flux measurements obtained with \ROSAT\ and \XTE. For \ROSAT\ we considered the CRs which were obtained during the \textit{All Sky Survey} \citep{Voges+00} and in a subsequent, specific observation \citep{Motch+97}. In their case, we used the \textit{WebPIMMS} tool\footnote{http://heasarc.gsfc.nasa.gov/Tools/w3pimms\_pro.html} and assumed our best--fit \textit{PL+BB} emission model to infer the source flux in the 2--10 keV energy band: in this way we estimated a 2--10 keV luminosity of, respectively, $(1.1\pm0.3) \times 10^{34}$ and $(2.4\pm0.2) \times 10^{34}$ erg s$^{-1}$. For \XTE\ we considered the value obtained with the two--blackbody model used by \citet{Reig&Roche99}, i.e. $\sim 2 \times 10^{34}$ erg s$^{-1}$. Although \RX\ was observed by \XMM\ at a larger flux level, all these luminosity estimates are in the range $10^{34}-10^{35}$ erg s$^{-1}$: the variation is smaller than a factor 10, therefore also the \XMM\ observation suggests that \RX\ is a persistent BeXRB. On the other hand, recently a large flux increase has been reported in three different events \citep{Morii+10,Krivonos+10b,Tsygankov+11}; based on the measured CRs, we have estimated that in all cases the source luminosity has risen up to a few $10^{36}$ erg s$^{-1}$. This implies a source variability of at least two order of magnitudes, in contrast with the classification of \RX\ as a persistent source. We analyzed also the \XTE\ ASM light--curve of \RX\ since 1996, in order to check the occurrence of other flux increases, but we found no clear evidence of such type of events in addition to the previous three ones; therefore the flux level of \RX\ is usually consistent with a persistent nature of the source, while the transient behaviour recently observed is likely due to structural changes in the circumstellar disc of the Be star.

From the spectral analysis, we obtained a hydrogen column density $N_{\rm H}$ = (7.3 $\pm$ 0.4)$\times10^{21}$ cm$^{-2}$, which is much lower (almost one order of magnitude) than the values estimated by \XTE. However, we note that the energy range of the \XTE\ spectral analysis (above 3 keV) is not well--suited for a good estimate of $N_{\rm H}$, while the low--energy end of the \XMM\ spectra (0.3 keV) allows a much more reliable analysis. LS V +44 17, the optical counterpart of \RX, has a color excess \textit{E(B-V)} = 0.65 \citep{Reig11}. Assuming $A_{\rm V}$ = 3.1 \textit{E(B-V)} and the average relation $A_{\rm V}$ = $N_{\rm H} \times 5.59 \times 10^{-22}$ cm$^{-2}$ between optical extinction and X--ray absorption \citep{PredehlSchmitt95}, this would give $N_{\rm H} = 3.6\times10^{21}$ cm$^{-2}$, a value which is a factor $\sim$ 2 lower than our result. We note that the data used to calibrate the previous relation is characterized by a comparable dispersion, therefore there is a rough agreement between the two measurements.

We found evidence of a previously undetected thermal component, in addition to the main power--law: it is characterized by a high temperature ($kT$ = 1.34 keV) and a small emission area ($R$ = 273 m), and contributes for $\sim$ 35 \% of the source flux below 10 keV. On the other hand, it was not possible to fit the \XMM\ spectra with the two--blackbody model used by \citet{Reig&Roche99}. The \XTE\ spectrum showed a possible feature at $\simeq$ 6.2 keV, suggesting the presence of a Fe--K emission line; but it was not possible to constrain its parameters, and only an upper limit of $\sim$ 100 eV on the Equivalent Width (EQW) could be provided. In spite of the higher luminosity level, we found no evidence of the Fe--K emission line and we set an upper limit of $\sim$ 70 eV (at 90 \% c.l.) to its EQW.

\begin{figure}[h]
\centering
\resizebox{\hsize}{!}{\includegraphics[angle=-90,clip=true]{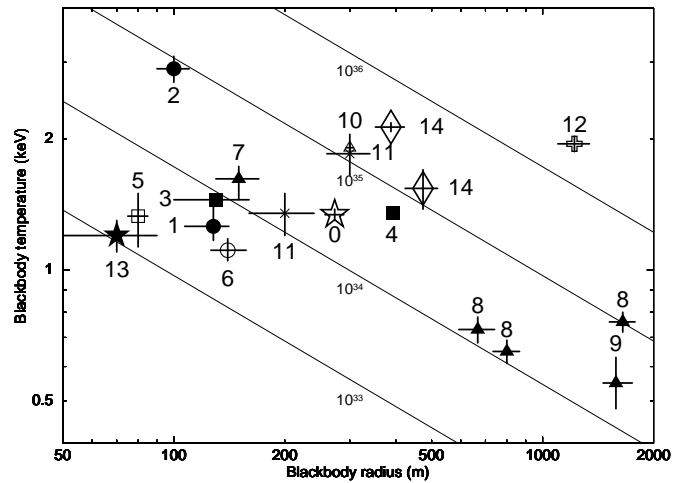}}
\caption{Best--fit values for radius and temperature of the \textit{BB} component in the case of \RX\ (\textit{empty star}), RX J1037.5-5647 (\textit{filled circles}), 4U 0352+309 (\textit{filled squares}), RX J0146.9+6121 (\textit{empty circle}), \TreA\ (\textit{empty square}), SAX J2103.5+4545 (\textit{empty triangle}), 4U 2206+54 (\textit{filled triangles}), IGR J11215-5292 (\textit{asterisks}), IGR J08408-4503 (\textit{cross}), XTE J1739-302 (\textit{filled star}), and SXP 1062 (\textit{empty diamonds}). The continuous lines connect the blackbody parameters corresponding to four different levels of luminosity (in erg s$^{-1}$). References: 0 - this work; 1- \citet{LaPalombara+09}; 2 - \citet{Reig&Roche99}; 3 - \citet{Coburn+01}; 4 - \citet{LaPalombara&Mereghetti07}; 5 - \citet{MukherjeePaul05}; 6 - \citet{LaPalombara&Mereghetti06}; 7 - \citet{Masetti+04}; 8 - \citet{Torrejon+04}; 9 - \citet{Reig+09}; 10 - \citet{Inam+04}; 11 - \citet{Sidoli+07}; 12 - \citet{Sidoli+09}; 13 - \citet{Bozzo+10}; 14 - \citet{Henault-Brunet+11}.}\label{BBparameters}
\end{figure}

For its size and temperature the \textit{BB} excess observed in \RX\ is similar to those detected in the other persistent Be/NS XBPs, i.e. LS I +61$^{\circ}$ 235/RX J0146.9+6121 \citep{LaPalombara&Mereghetti06}, X Persei/4U 0352+309 \citep{Coburn+01,LaPalombara&Mereghetti07} and LS 1698/RX J1037.5-5647 \citep{Reig&Roche99,LaPalombara+09}: therefore this \textit{hot BB} spectral component is a common property of this type of sources. However, we note that a similar feature has been observed also in \TreA\ \citep{Orlandini+04,MukherjeePaul05}, 4U 2206+54 \citep{Masetti+04,Torrejon+04,Reig+09}, and SAX J2103.5+4545 \citep{Inam+04}, which are other types of XBPs with low luminosity ($L_{\rm X}\le10^{35}$ erg s$^{-1}$) and long pulse period ($P >$  100 s). Moreover, recently the same type of \textit{BB} excess has been detected also in the three Supergiant Fast X--ray Transients (SFXTs) IGR J11215-5292 \citep{Sidoli+07}, IGR J08408-4503 \citep{Sidoli+09} and XTE J1739-302 \citep{Bozzo+10}, only the first of which is a confirmed pulsar ($P$ = 187 s). Finally, very recently a hot \textit{BB} excess has been detected also in the SMC binary pulsar SXP 1062, a possible new persistent Be X--ray binary \citep{Henault-Brunet+11}. In Fig.~\ref{BBparameters} we report the best--fit radius and temperature for the \textit{BB} component of these sources, together with lines showing four different levels of the blackbody luminosity; for some source more than one set of values is shown, corresponding to different observations or flux levels. In most cases the spectral parameters are within a narrow range of values, i.e. $kT_{\rm BB} \sim$ 1--2 keV and $R_{\rm BB} <$ 200 m: we emphasize that, in all these cases, the estimated total source X--ray luminosity is $\sim 10^{34}$ erg s$^{-1}$, with a 20--40 \% contribution of the blackbody component. For \RX\ the \textit{BB} radius is slightly higher ($R_{\rm BB} \sim$ 270 m), in agreement with the fact that its total X--ray luminosity is also higher ($L_{\rm X} \sim 8\times10^{34}$ erg s$^{-1}$): when observed at high (i.e. $L_{\rm X} \sim 10^{35}$ erg s$^{-1}$) luminosity level, also RX J1037.5-5647 (point 2), 4U 0352+309 (point 4), SAX J2103.5+4545 (point 10) and IGR J11215-5292 (point 11) show large values of temperature and/or radius.

\begin{figure}[h]
\centering
\resizebox{\hsize}{!}{\includegraphics[angle=-90,clip=true]{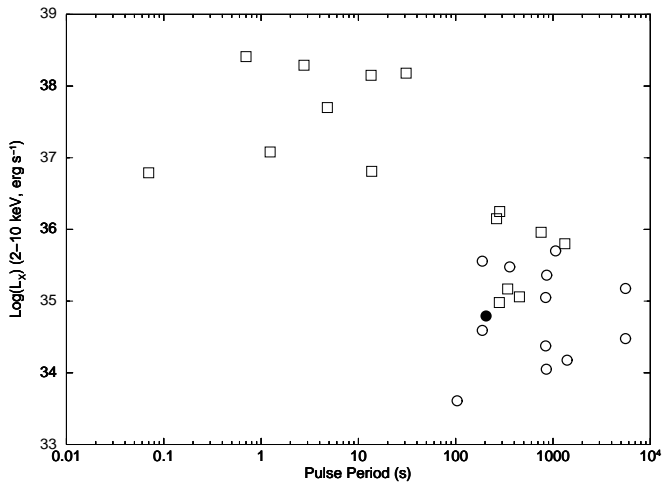}}
\caption{X--ray luminosity (in the 2--10 keV energy range) of the XBPs with a detected thermal excess as a function of the pulse period. The \textit{filled circle} refers to the \XMM\ observation of \RX, \textit{empty circles} refer to other detections of \textit{hot BB} pulsars, \textit{open squares} to the \textit{soft excess} sources.}
\label{luminosity_period}
\end{figure}

In contrast to this sample of sources, several XBPs are characterized by a \textit{soft} excess, since the fit of this component with a thermal emission model provides low temperatures ($kT <$ 0.5 keV) and large emitting regions (\textit{R} $>$ 100 km). In Fig.~\ref{luminosity_period} we report the luminosity and pulse period of both types of XBPs: the \textit{soft excess} and the \textit{hot BB} ones are reported as \textit{squares} and \textit{circles}, respectively. Based on their distribution in the $P - L_{\rm X}$ diagram, these pulsars are divided into two distinct groups: the sources in the first group are characterized by high luminosity ($L_{\rm X}\ge10^{37}$ erg s$^{-1}$) and short pulse period (\textit{P} $<$ 100 s), and in most cases they are in close binary systems with an accretion disk; those in the second group have low luminosities ($L_{\rm X}\le10^{36}$ erg s$^{-1}$) and long pulse periods (\textit{P} $>$ 100 s), since they have wide orbits and are wind--fed systems. While all the pulsars in the first group are characterized by a \textit{soft excess}, both types of pulsars are present in the second group. In this case, the \textit{hot BB} pulsars are the ones that, on the average, are characterized by the lowest luminosities and the longest periods. This suggests that the \textit{hot BB} spectral component is a common feature of the low--luminosity and long--period XBPs. However, in the second group of sources there is no clear separation between the two types of pulsars, since there is a partial overlap between the \textit{soft excess} and the \textit{hot BB} ones. Therefore, the origin of the spectral difference in pulsars having comparable values of luminosity and pulse period is unclear. Even if it could be explained by really different emission mechanisms at work in these sources, we can not exclude that, at least in some cases, it is due to the degeneracy of spectral fitting in some spectra. In fact, on the basis of the X--ray spectrum it is often impossible to distinguish between a \textit{BB} with small radius and high temperature and one with large radius and low temperature \citep[see, e.g.,][]{Bozzo+10}.

Based on our results and on the emission models proposed by \citet{Hickox+04}, also in the case of \RX\ it makes sense to attribute the observed spectral excess to the thermal emission from the NS polar cap, in agreement to what already suggested for the other persistent Be pulsars. In order to check this hypothesis, we assume that the source is in the `accretor' status, with matter accretion on the NS surface. Assuming $M_{\rm NS}$ = 1.4 $M_{\odot}$ and $R_{\rm NS}$ = $10^6$ cm, the source luminosity $L_{\rm X} = 8.3 \times 10^{34}$ erg s$^{-1}$ implies an accretion rate $\dot M \simeq 4.5 \times 10^{14}$ g s$^{-1}$ and, adopting $B_{\rm NS} = 10^{12}$ G, a magnetospheric radius $R_m \simeq 10^9$ cm \citep{Campana+98}. In this case, based on the relation $R_{col} \sim R_{\rm NS}$ ($R_{\rm NS}/R_m$)$^{0.5}$ \citep{Hickox+04}, we would obtain $R_{col} \sim$ 320 m. This value, which is an estimate of the expected size of the polar cap, is remarkably near to the estimated blackbody emitting radius ($R_{\rm BB} \sim$ 270 m): therefore it strongly supports the suggestion of a polar--cap origin for the observed \textit{BB} emission. If this description is correct, we would expect to observe some variability of the thermal component along the pulse phase. The phase--resolved spectral analysis confirmed the spectral variability along the pulse, but it can be attributed to the \textit{PL} component, since a constant \textit{BB} component is fully compatible with the spectral data. The lack of variability of the \textit{BB} component is not in contradiction with a polar--cap origin, as an anisotropic radiation can appear steady along the pulse for some geometrical configurations of the hot spots on the NS surface \citep{Beloborodov02}. On the other hand, the phase dependence of the non--thermal component is likely due to the anisotropy of the Comptonized radiation of the accretion column. This is shown, for example, in the self--consistent model by \citet{Becker&Wolff07} for the dynamics and the radiative transfer occurring in the accretion column of bright X--ray pulsars. This model is based on the physical picture originally proposed by \citet{Davidson73}, in which the accreting gas passes through a radiative, radiation--dominated shock before settling onto the NS surface. For the relatively low luminosity of \RX\ the shock could be absent or have only a limited extent above the neutron star surface. Thus the blackbody emission from the dense thermal mound at the base of the column could be the dominant seed component for the Comptonization. In this geometry, the radiation is probably emerging as a `pencil beam' rather than from the column's lateral surface as in the Becker \& Wolff model.

\section{Conclusions}

We have analysed a $\sim$ 17 ks \XMM\ observation of the Be/NS X--ray pulsar \RX. The source was detected at a luminosity level $L_{\rm X}\simeq 8.3\times10^{34}$ erg s$^{-1}$ in the 0.3--12 keV energy range: this value is less than one order of magnitude higher than the average level of the previous \ROSAT\ and \XTE\ observations, therefore it confirms the persistent nature of this source. This classification is not denied by the large flux increases recently observed, which imply luminosity levels of a few $10^{36}$ erg s$^{-1}$, since they can be explained by structural changes in the circumstellar disc of the Be mass donor.

The \XMM\ observation has provided a refined pulse period $P$ = 204.96 $\pm$ 0.02 s, which, compared to previous measurements, implies an average pulsar spin--down $\dot P = (6.4\pm1.3)\times10^{-9}$ s s$^{-1}$ between 1998 and 2011. It implies a low average accretion rate, in agreement with the persistent nature of the pulsar. The pulse profile shows a complex structure, which is not sinusoidal, and at all energies the PF is $\sim$ 55 \%. However the HR is characterized by significant variations, expecially around the pulse minimum, which indicates a spectral variability of the source.

The source spectrum shows a count excess above the main power--law component, described by a blackbody with $kT_{\rm BB} \simeq$ 1.3 keV and $R_{\rm BB} \simeq$ 270 m, which contributes $\sim$ 35 \% to the source luminosity between 0.3 and 12 keV. We found no evidence of a narrow iron K$_{\alpha}$ line between 6 and 7 keV, with an upper limit of $\sim$ 70 eV on its equivalent width. The blackbody radius is comparable to the estimated size of the NS polar--cap, which suggests that the origin of this component is on the NS surface. The phase--resolved spectroscopy neither confirms nor disproves this scenario, since the observed spectral variability along the pulse period can be attributed to the power--law component and is consistent with a constant thermal component.

The spectral properties of \RX\ are in full agreement with those observed in the other three persistent Be binary pulsars 4U 0352+309, RX J0146.9+6121 and RX J1037.5--5647, thus confirming that the \textit{hot BB} spectral component is a common property of this type of sources. Moreover, we have shown that the same type of feature has been detected also in other low--luminosity and long--period pulsars; therefore it is an ubiquitous phenomenon which requires further investigations.

\begin{acknowledgements}
This work is based on observations obtained with \XMM, an ESA science mission with instruments and contributions directly funded by ESA Member States and NASA. We acknowledge financial contributions by the Italian Space Agency through ASI/INAF agreements I/009/10/0 and I/032/10/0 for, respectively, the data analysis and the \XMM\ operations. PE acknowledges financial support from the Autonomous Region of Sardinia through a research grant under the program PO Sardegna FSE 2007--2013, L.R. 7/2007 `Promoting scientific research and innovation technology in Sardinia'.
\end{acknowledgements}

\bibliographystyle{aa}
\bibliography{biblio}

\end{document}